# Magnetic Fields on the Flare Star Trappist-1: Consequences for Radius Inflation and Planetary Habitability

D. J. Mullan[1], J. MacDonald[1], S. Dieterich[2], and H. Fausey[1]

[1]Dept. of Physics and Astronomy, University of Delaware, Newark DE 19716
[2]Dept. Terrestrial Magnetism, Carnegie Institution of Washington, Washington, DC 20015, U.S.A

**Abstract**
We construct evolutionary models of Trappist-1 in which magnetic fields impede the onset of convection according to a physics-based criterion. In the models that best fit all observational constraints, the photospheric fields in Tr-1 are found to be in the range 1450-1700 G. These are weaker by a factor of about 2 than the fields we obtained in previous magnetic models of two other cool dwarfs (GJ65A/B). Our results suggest that Tr-1 possesses a global poloidal field that is ~100 times stronger than the Sun's global field. In the context of exoplanets in orbit around Tr-1, the strong poloidal fields on the star may help to protect the planets from the potentially destructive effects of coronal mass ejections. This, in combination with previous arguments about beneficial effects of flare photons in ultraviolet and visible portions of the spectrum, suggests that conditions on Tr-1 are not necessarily harmful to life on a planet in the habitable zone of Tr-1.

1. Introduction

Trappist 1 (Tr-1) is an M8 dwarf star with an estimated age reported by Burgasser and Mamajek (2017: hereafter BM17) to be 7.6 ± 2.2 Gyr. The star is surrounded by at least seven exoplanets, of which 3 are believed to lie in the habitable zone (HZ), and are therefore of interest in the context of extraterrestrial life. Moreover, with an age which (according to BM17) is at least as old as the Sun, there has been at least as much time for life to emerge on a Tr-1 planet as in the case of the Earth. We shall return to a discussion of the age of Tr-1 in the context of rotation and spot coverage in Section 2 below.

According to BM17, the stellar radius (0.121 ± 0.003 $R_\odot$) is inflated relative to the predictions of solar-metallicity evolutionary models for a star with mass 0.08 $M_\odot$ by 8-14%. However, these percentage inflations in the radius depend sensitively on the value one chooses for the stellar mass: Van Grootel et al. (2018) conclude that the mass of Tr-1 is not as small as BM17 suggest. Instead, Van Grootel et al (2018) cite a mass of 0.089 ± 0.006 $M_\odot$: in such a case, the percentage radial inflation, while still present, would not be as large as the values reported by BM17. The occurrence of an empirically inflated radius on Tr-1 is the principal reason why we undertake the magnetic modeling of Tr-1 described in Section 3 below.

Vida et al. (2017) have reported on a study of 42 flares on Tr-1 detected by the Kepler spacecraft. Because of the possible dangers to living organisms which might be posed by the effects of flares on exoplanets in orbit around Tr-1, Vida et al. state (in their abstract) that the flares "make these [exoplanets] less favorable for hosting life". And in the title of their paper, Vida et al (2017) explicitly include the question "Unsuited for Life?"

In the present paper, our goal is to re-examine this question of suitability for life on a Tr-1 planet using information we obtain from models of the star that include the structural effects of magnetic fields.

In Section 2, we summarize the information that can be extracted from photometric data concerning the properties of magnetic features on the surface of Tr-1. In Section 3, we present the principal results of this paper, namely, quantifying the internal structure of Tr-1 in the context of a magneto-convective model: this model leads to the key physical quantity of interest in the present paper, i.e. the strength of the magnetic field on the surface of Tr-1. In Section 4, we discuss our estimates of the surface magnetic field in terms of the global components of the field on Tr-1. In Section 5, we consider some quantitative details as to how the surface magnetic fields on Tr-1 contribute to flaring activity on that star. In Section 6, we consider the appropriateness of making comparisons between the magnetic fields in stars that differ in internal structure. In Sections 7 and 8, we re-examine the question of suitability for life on a Tr-1 planet from two points of view: (i) mass ejections (Section 7) and (ii) photons (Section 8). Conclusions are discussed in Section 9.

## 2. Rotational period and variability: spots and faculae

The rotational period of Tr-1 is variously reported as 3.3 days (Luger et al. 2017; Vida et al. 2017), 1.40 days (Gillon et al. 2016), and 0.819 days (Roettenbacher & Kane 2017). These rotation periods are shorter than the Sun's by factors of 10-30, i.e. the angular velocity $\Omega$ in Tr-1 exceeds $\Omega$(Sun) by 10-30. Kippenhahn (1973) suggests that the strength of a dynamo field B should increase as $\Omega$ increases according to $B \sim \Omega^x$, where x = 1 or 1.5. If such scalings are relevant to Tr-1, then the dynamo field strength on Tr-1 could exceed that in the Sun by factors that are at least as large as 10-30, and perhaps as large as 30-160. In the present paper, we examine this possibility quantitatively from the perspective of stellar structure by constructing magneto-convective models of Tr-1.

It is important to note that Luger et al. (2017), using an 80-day segment of data from the Kepler K2 mission, report a light curve for Tr-1 that suggests (to the eye) that the most prominent photometric features appear to be excursions away from the overall average brightness towards *larger* brightnesses. And the amplitudes of the excursions are on average seen (by visual inspection) to be at the ~1% level. This is in contrast to other cool stars in two respects: (i) qualitatively, in the presence of cool starspots, the most prominent features in the light curve are well-defined excursions to *smaller* brightnesses; (ii) quantitatively, the amplitudes of the excursions on spotted stars can be as large as 30-40% (e.g. Bopp & Evans 1973; Strassmeier, 1999). In the Sun, excursions to larger brightnesses are associated with localized bright features labelled photospheric faculae (e.g. Bray & Loughhead 1964: hereafter BL64). Rackham et al. (2018) have quantified the separate contributions of spots and faculae to the photometric variations in Tr-1: they conclude that spots occupy 8% of the surface area, while faculae occupy 54%. The analysis of Rackham et al. assumed that the spots and faculae are long-lived features. Relaxing the assumption of longevity, Morris et al. (2018) analyzed the Tr-1 variability in terms of shorter-lived surface features that they called "bright starspots". Morris et al. suggest that the bright starspots have radii $R_{spot}$ of about 0.004 $R_\star$: with $R_\star = (0.121 \pm 0.003) R_\odot = 8.5 \times 10^9$ cm, we find that the spots have linear radii of about 300 km. Such radii are reminiscent of facular elements on the Sun, which have linear dimensions of < 300 km and are magnetic flux tubes located in the dark intergranular lanes. In the Sun, granules have horizontal dimensions of order 1000 km, and the dark intergranular lanes are smaller than this by a factor of a few. In view of this, we suggest that the "bright starspots" mentioned by Morris et al. may be considered as analogs of faculae, i.e. individual magnetic flux ropes in which the reduced density allows one, when viewing at a finite angle from the vertical, to see downwards to the hot walls of the flux rope in deeper layers (Spruit 1976).

In fact, in the Sun it is observed that the radiant flux varies during the solar cycle in such a way that when sunspots are most abundant, the solar luminosity is *larger* than average (e.g. Radick et al. 2018). This indicates that the flux excesses from faculae in the Sun more than compensate for the flux deficits in sunspots. Thus, the Sun is a star where the temporal variability associated with the activity cycle is dominated by faculae. Radick et al. (1998) reported that, depending on the intensity of chromospheric activity, there is a transition from facular-dominated variability to spot-dominated variability. The sense of the transition is that stars where the activity level is lower tend to exhibit facular-dominated variability (such as the Sun), whereas more active stars have variability that is dominated by spots. These conclusions of Radick et al. (1998) were reinforced by Shapiro et al. (2014), who also extended the properties to include effects of different angles between the rotation axis and the line of sight.

Given the predominance of facular variability in Tr-1, it seems plausible to conclude that Tr-1 resembles the Sun to the extent that both can be assigned to the "low-activity set" of magnetically active stars.

On the other hand, the "low-activity" label for Tr-1 has recently been challenged by Dmitrienko and Savanov (2018: hereafter DS). Using the same Kepler K2 data set as Luger et al (2017), but reduced using different algorithms, DS found that the amplitude of Tr-1 light excursions was not 1% (as estimated above), but 2.6-3.1%. Interpreting these fluctuations in terms of spots with certain temperatures (160 K cooler than the photosphere), DS calculated that a fractional area $S = 5$-6% of the surface of Tr-1 is covered by spots. By comparing with their previous estimates of $S$ values in a sample of 1570 M dwarfs with known ages, DS claimed that Tr-1 is a high-activity star with an "$S$-age" of less than 200 Myr. Such a young age is problematic for stellar evolution theory: as will be seen in Fig. 1 below, a star with an age of 200 Myr and a mass of 0.08-0.09 $M_\odot$ would still be in its *pre*-main sequence phase of evolution, with properties which would deviate by many σ from the empirical data for Tr-1. Moreover, if the age is really as low as 200 Myr, the interest in Tr-1 as regards the emergence of life would diminish considerably: on Earth, it is estimated that some 200 Myr had to elapse before the first life emerged (e.g. **Dodd et al 2017**). However, we note that in Fig. 4 of DS, there is considerable scatter in the $S$-age data: in fact, within the error bars plotted by DS for Tr-1, there are several M dwarfs with $S$-ages of more than 1 Gyr: the two oldest of these have $S$-ages of 2.6-3.1 Gyr. Furthermore, DS also refer to X-ray data on Tr-1 which suggest an age of 2±1 Gyr: DS do not explain how to reconcile the 10-fold discrepancy between X-ray age and their claim of 0.2 Gyr for the *S*-age. In the context of the evolutionary calculations to be reported in the present paper, we note that stars with masses in the range 0.08-0.09 $M_\odot$ reach the main sequence in times of 0.5-0.8 Gyr. Once on the main sequence, their lifetimes ($\sim M/L \sim 1/M^{2-3}$) are $\geq 10^3$ Gyr. With such long lifetimes, the differences in *L* and *R* between a star with an age of 2-3 Gyr and a star with an age of 5-7 Gyr are negligible on the scales of our Figs. 1-5. Hence, our results can be applied without significant error to empirical data whether the "true" age of Tr-1 has the value reported by BM17 or the X-ray age reported by DS.

## 3. Evolutionary models of Trappist-1: derivation of surface magnetic field strength

The possibility that Tr-1 has an empirical radius that exceeds the radius of a standard (non-magnetic) model of a star with a specified mass suggests that it is worthwhile to explore the case of a stellar model in which a global magnetic field is present. Such a model can be derived based on a quantitative formulation of the physics involved in a low-mass star where the presence of a vertical magnetic field in an electrically conducting medium impedes the onset of convection (Mullan & MacDonald 2001; MacDonald & Mullan 2012, 2014, 2017a).

*3.1. The observational data*

In this sub-section, we present the constraints on stellar modelling which are set solely by observational data: i.e. we do not consider in this sub-section any results that depend on stellar models.

The metallicity has been determined from near-infrared spectroscopy (Gillon et al. 2016) to be near solar, [Fe/H] = +0.04 ± 0.08. From a comprehensive analysis of empirical age constraints, BM17 determined that Tr-1 has an age of 7.6 ± 2.2 Gyr. From analysis of the transit light curves, Delrez et al. (2018) determine the mean stellar density to be $\rho = 51.1^{+1.2}_{-2.4}$ in units of the mean density of the Sun. From analysis of the optical data for Tr-1, Van Grootel et al. (2018) determined the parallax to be $82.4 \pm 0.8$ mas, i.e. the distance to Tr-1 is $12.14 \pm 0.12$ pc. This distance is considerably more precise than the distance obtained by Costa et al. (2006), although the mean values of the distance overlap within the error bars. Using the spectral energy distribution of Filippazzo et al. (2015), Van Grootel et al. (2018) find the luminosity of Tr-1 to be $L = (5.22 \pm 0.19) \times 10^{-4} L_\odot$.

This set of data is insufficient by itself to determine the stellar mass or stellar radius. We, therefore, have applied the method of Dieterich et al. (2014) (also see MacDonald et al. 2018) to determine the luminosity and effective temperature of Tr-1 from published photometry (Filippazzo et al. 2015). We find, based on the parallax measurement of Van Grootel et al., that $L = (5.14 \pm 0.14) \times 10^{-4} L_\odot$ and $T_{eff} = 2519 \pm 51$ K. The corresponding radius value is $R = 0.1190 \pm 0.0051 R_\odot$. Using this radius with the mean density measurement from Delrez et al., we find the stellar mass is $M = 0.0858 \pm 0.0114 M_\odot$.

From analysis of the FeH absorption band, Reiners & Basri (2010) determined that Tr-1 has a magnetic field averaged over the visible surface of $Bf = 600^{+200}_{-400}$ G.

BM17 use a number of empirical age constraints from kinematics, rotation, magnetic activity, lithium absorption, position in the color-absolute magnitude diagram, average density, surface gravity features, and metallicity to determine that Tr-1 has an age of 7.6 ± 2.2 Gyr.

*3.2. Previous modelling*

Using only the mean stellar density and their luminosity value as constraints, Van Grootel et al. (2018) determined from their stellar modeling that the stellar mass would need to be 0.081 ± 0.003 $M_\odot$ and the age would then be 450 ± 55 Myr. The corresponding model radius and effective temperature are 0.117 ± 0.002 $R_\odot$ and 2555 ± 25 K. Because this age is in clear conflict with that found by BM17, Van Grootel et al. also applied only the luminosity and age constraints to their stellar evolution modeling. In this case, agreement is found for a stellar mass of 0.089 ± 0.003 $M_\odot$ and an age between 2 and 15 Gyr. The corresponding model radius and effective temperature are 0.114 ± 0.002 $R_\odot$ and 2595 ± 30 K. No reasonable fit was found at solar metallicity when the constraints from luminosity, age, and density are applied together. The stellar density at ~0.09 $M_\odot$ is found to be much higher than the value measured from transits because the model stellar radius is too low. Van Grootel et al. point to 'the usual suspects' for this radius anomaly: 1) the presence of strong magnetic field causing the stars to inflate by inhibiting convective energy transport (Mullan & MacDonald 2001) and/or by creating surface spots (Chabrier et al. 2007), or 2) higher than solar metallicity causing the star to inflate due to larger opacity (Feiden & Chaboyer 2014).

In the present paper, we report on a quantitative investigation that seeks to interpret the empirical inflation of the radius of Tr-1 in the context of magnetic inhibition of the onset of convective energy transport.

*3.3. Magneto-convective modeling*

Although our model for magneto-convection was first described in Mullan & MacDonald (2001), improvements have been added to the model in the intervening years. A description of the current model is provided by MacDonald et al. (2018). In the latter paper, we compared the theoretical magnetic field strengths required to give the observed radius inflations of the components of the M-dwarf binary GJ 65 with the field strengths determined by Kochukhov & Lavail (2017) and Shulyak et al. (2017). Other aspects of our stellar evolution code are described in MacDonald & Gizis (2018).

We have made one important modification to the method described in MacDonald et al. (2018) for construction of magnetic stellar models for Tr-1. As a first step in our method, we construct non-magnetic models using boundary conditions from atmosphere calculations based on the $T - \tau$ relation of Krishna-Swamy (1966: hereafter KS): our goal is to have KS models match the photospheric properties of non-magnetic models that use boundary conditions from BT-Settl atmosphere models (Allard et al. 2012a, 2012b; Rajpurohit et al. 2013). This requires us to adjust the mixing length parameter, α, in the KS atmosphere calculations. In MacDonald et al. (2018), for reasons of simplicity, a single value for α appropriate for low mass main sequence models, α = 0.7, was used. In the present paper, we use the value α of that gives the best match to the BT-Settl boundary conditions for given log $g$ and $T_{eff}$.

Our adopted magnetic field profile depends on two-parameters: A magnetic inhibition parameter, $\delta$ and a magnetic field ceiling $B_{ceil}$. The magnetic inhibition parameter, which was first introduced by Gough & Tayler (1966: hereafter GT), is a local parameter defined by

$$\delta = \frac{B_v^2}{4\pi\gamma P_{gas} + B_v^2}, \qquad (1)$$

where $B_v$ is the vertical component of the magnetic field, $P_{gas}$ is the gas pressure, and γ is the ratio of specific heats. In the original GT criterion, convective stability is ensured as long as the radiative gradient $\nabla_{rad}$ does not exceed $\nabla_{ad} + \delta$, where $\nabla_{ad}$ is the adiabatic gradient. In the original GT paper, the primary interest was in the conditions which exist *in the Sun*: in such an environment, the temperatures of the gas in which magnetic fields and convection interact are high enough that the electrical conductivity has a large value which can (without significant error) be considered to be in effect infinite in the context of the ambient convective cells. However, in stars which are clearly cooler than the Sun (such as GJ65 A/B and Tr-1), the assumption of infinite electrical conductivity is no longer valid. At temperatures of 2600 K or less, the question which plays a key role in the field-gas interaction is the following: in the course of the (finite) time period required for material in a convective cell to complete one circulation of the cell, how effectively can the magnetic fields become "unfrozen", and "slip" through the gas? In cases where the "slippage" is large enough to amount to a significant fraction of the linear size of a convective cell, a magnetic field of given strength is less effective at impeding the onset of convection than such a field would be in the Sun (MacDonald et al. 2018). To quantify the effects of finite conductivity, we replace the original GT criterion with a modified "GTC criterion" (MacDonald & Mullan 2009). Our GTC code includes the effects of field line "slippage" through the mostly neutral gas near the photosphere (MacDonald & Mullan 2017a), and allows for deviations from ideal gas behavior (Mullan & MacDonald 2010).

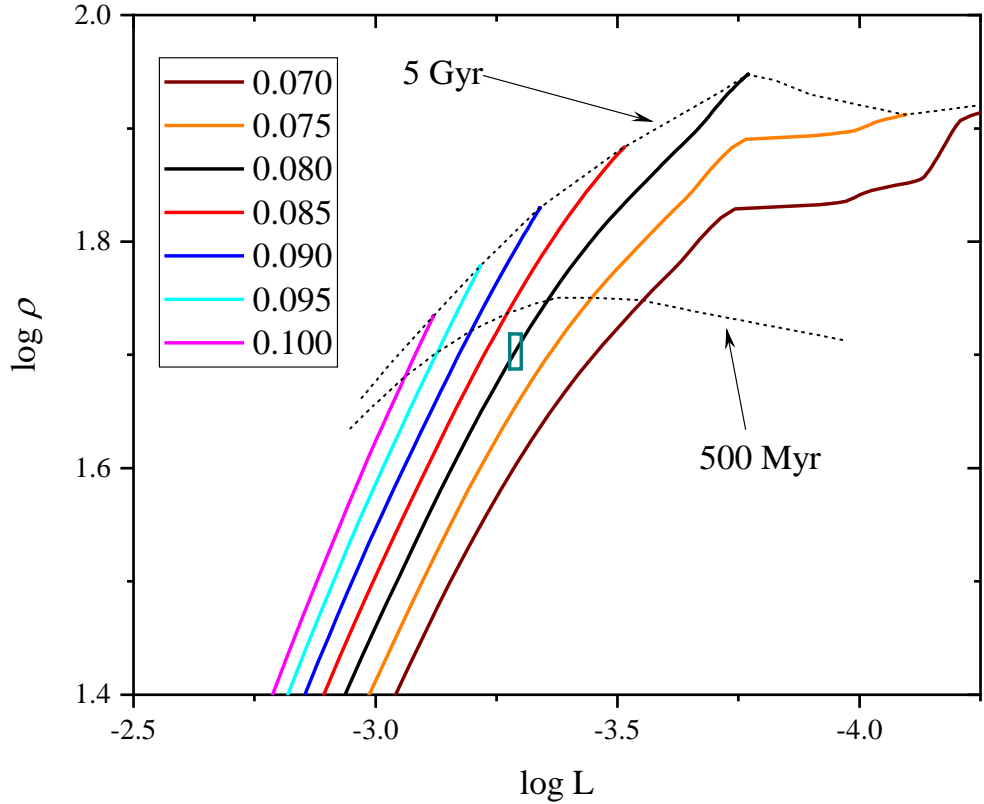

Figure 1. Evolutionary tracks of non-magnetic stellar models in the log(luminosity) – log(mean density) plane. The mass on a track of given color is listed in the insert (in units of $M_\odot$). The broken lines in the figure are isochrones for models of ages of 500 Myr and 5 Gyr. The empirical location of Tr-1 is shown by the dark cyan rectangle. Solar units are used for $L$ and $\rho$.

*3.4. Models that do not include magnetic effects*

Before considering models that include the effects of magnetic fields on stellar structure, we explore the possibility that the observed low density of Tr-1 could be the result of higher than solar heavy element abundances. We begin by presenting results for models that have the same abundances as the BT-Settl atmosphere models, i.e. the solar photospheric abundances derived by Caffau et al. (2011). The mass fractions of hydrogen, helium and heavy elements are $X = 0.7321$, $Y = 0.2526$, and $Z = 0.0153$, respectively. The models have masses that cover a range that is consistent with our temperature determination, i.e. 0.075 to 0.100 $M_\odot$, in increments of 0.005 $M_\odot$. We show in Figure 1 evolutionary tracks in the luminosity – mean density plane with the masses (in units of solar mass) shown in the legend. The broken lines in Figure 1 display isochrones for our models, one with an age of 500 Myr, and the other with an age of 5 Gyr. The empirical location of Tr-1 in this plane is shown by the small colored rectangle.

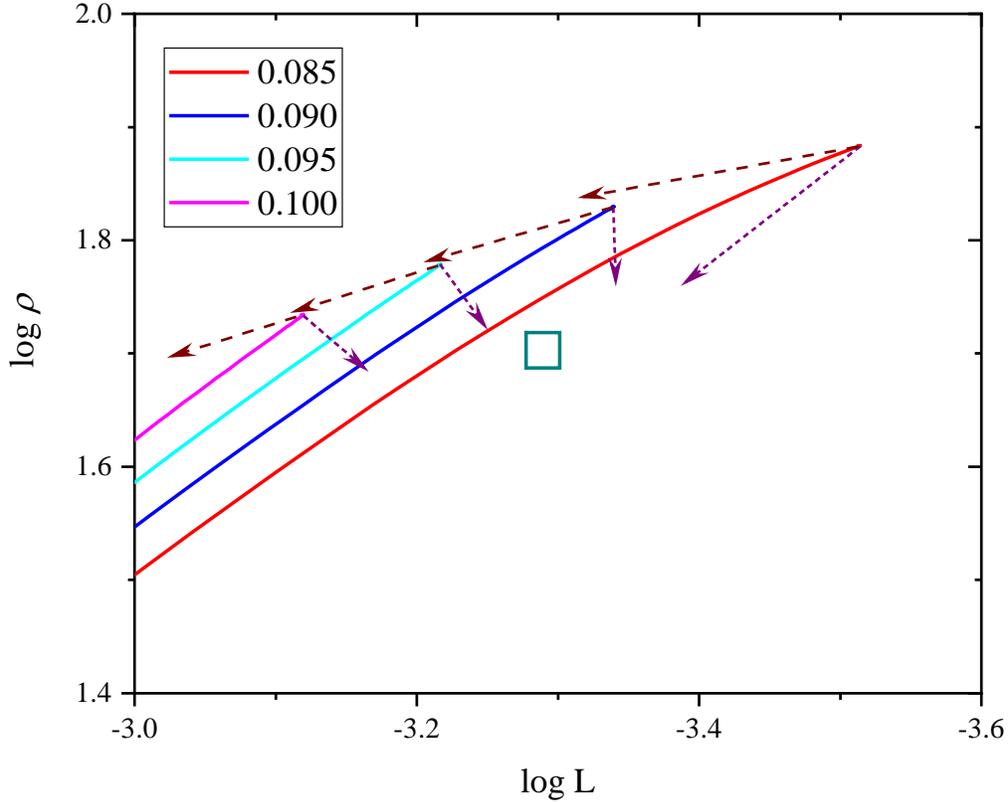

Figure 2. Evolutionary tracks of low-mass stars plotted in the log L – log $\rho$ plane. Notation is the same as in Fig. 1. The long-dashed brown and short-dashed purple arrows show the directions of the model main sequence shifts when $Y$ and $Z$ (respectively) are independently increased. The lengths of the arrows correspond to increments of log $Y$ and log $Z$ by 0.1 and 0.3, respectively. The empirical location of Tr-1 is shown by the dark cyan rectangle.

    The fact that, in Figure 1, the 0.080 $M_\odot$ track (the black line) passes through the empirical rectangle for Tr-1, at first sight might lead us to conclude that the mass quoted by BM17 (0.080 $M_\odot$) *could* be a good fit to the empirical data. That conclusion might be correct if we were to restrict our consideration to the empirical values of only *two* quantities ($L$ and $\rho$), and disregard other parameters. However, such a conclusion is ruled out when we also take into account another parameter, namely, the age of Tr-1. It is important to note that the empirical rectangle for Tr-1 in Figure 1 is situated in a position that lies definitely *below* the 500 Myr isochrone: such a position would require that Tr-1 have an age that is *less* than 500 Myr. That age would be much too young to be consistent with the age of Tr-1 as reported by BM17, a conclusion that was also reached by Van Grootel et al. (2018). **For completeness, we should mention that an age of less than 500 Myr *could* be consistent with the *S*-age quoted by DS: however, we have already mentioned (see Section 2 above) that the DS estimate of *S*-age could be too small by an order of magnitude, while a second age estimate by DS (based on X-rays) is also too long (by an order of magnitude) to be consistent with the quoted *S*-age. In view of these properties, we will in what follows focus on Tr-1 models which are on the main sequence.**

We next explore the possibility that the mass, age, luminosity and mean density constraints could be simultaneously satisfied by non-magnetic models in which the abundances of the elements take on different numerical values. Because the BT-Settl atmospheres (which do not incorporate magnetic fields) are available for only a few sets of abundances, we perform a differential analysis using models computed with a simple Eddington boundary condition. Figure 2 shows the effects of increasing $Y$ (alone) or $Z$ (alone) on the location of the model main sequence phase. Increasing $Y$ while keeping $Z$ fixed is equivalent to increasing the stellar mass while retaining $Y$ and $Z$ at their original values.

We see from Fig. 2 that there is a narrow range of mass, between 0.090 and 0.095 $M_\odot$, for which the main sequence *could* in principle be shifted to a location consistent with the empirical rectangle in the $L$-$\rho$ plane by simply increasing $Z$ while keeping $Y$ fixed. In fact, if we were permitted to increase $Z$ to any value whatsoever, no matter how large, the empirical rectangle could likely be replicated. However, the question which is of quantitative relevance here is: how much would $Z$ have to be increased in order to obtain consistency with the empirical rectangle? The answer is found to be as follows: the minimum value of $Z$ for which our models would agree with the empirical rectangle at the 1-$\sigma$ level is $Z = 0.036$. The corresponding $Z/X$ ratio is 0.051. In order to determine the correspondence with the observed [Fe/H] = $+0.04 \pm 0.08$ (Gillon et al 2016) requires knowledge of the solar $Z/X$ ratio: this remains a hot topic of debate, with values ranging from 0.0165 (Asplund, Grevesse, & Sauval 2006) to 0.0274 (Anders & Grevesse 1989) and a helioseismologically preferred value of 0.0233 (Antia & Basu 2006). A value of $Z/X$ = 0.051 corresponds to [Fe/H] = 0.27 – 0.49 which is discrepant by 2.9 – 5.6 $\sigma$ from the mean value reported by Gillon et al. (2016). These discrepancies are so large that we conclude that the radius inflation observed in Tr-1 is unlikely to be solely due to high metallicity.

The conclusion that metal abundances alone cannot explain the radius inflation in low-mass stars which are metal poor has already been discussed in earlier work on the metal-poor binary CM Dra (MacDonald & Mullan 2012): we were able to strengthen this conclusion by comparing CM Dra to a binary which has similar masses as CM Dra, but with metal-*rich* composition (KOI 126). The contrast was striking: in KOI 126, the inflated radii were found to agree with a model with enhanced metals alone, and there was no need to call upon magnetic effects to fit the data. But for CM Dra, with its low metal abundance, we could not explain the inflated radii on the basis of enhanced metals: magnetic fields were, however, found to be adequate to the task.

*3.5. Results of magneto-convective modeling of Tr-1*

Now we turn to models in which the effects of magnetic fields are explicitly included in the criterion for the onset of convection. As shown by Mullan & MacDonald (2001), two principal characteristics emerge from magneto-convective models: (i) the radius of a magnetic model with a given mass is *larger* than that of a non-magnetic model with the same mass, (or equivalently the mean density of a magnetic model is smaller than a non-magnetic model with the same mass); (ii) the effective temperature of the magnetic model is *smaller* than that of the non-magnetic model with the same mass. In the process of computing our magneto-convective models, we vary the strength of the surface magnetic field in order to achieve the following:

given the ages which have been determined by BM17, what value of the field strength is required to give consistency not only with the mean density, but also with the empirical values of radius, effective temperature, and luminosity?

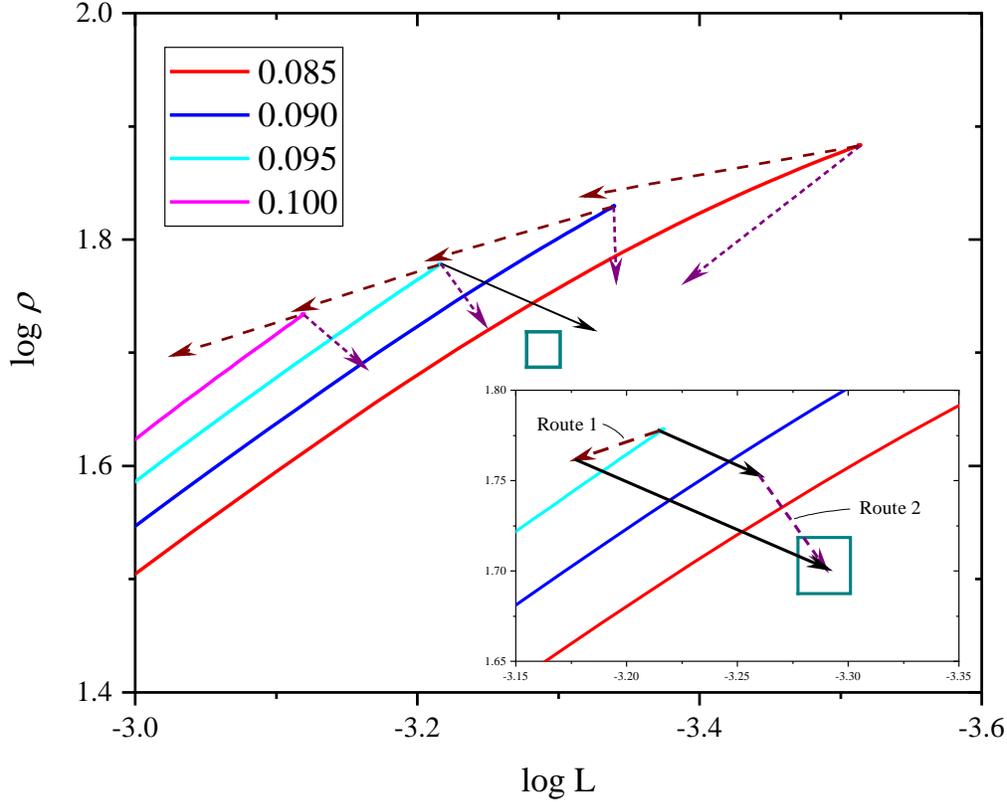

Figure 3. As Fig. 2, except for the addition of a thin solid black line with arrow in the main part of the figure: this shows the direction of shift in the location of the main sequence of a 0.095 $M_\odot$ model due solely to inclusion of the magnetic inhibition of convection in the absence of any changes in $Y$ or $Z$.

Figure 3 is similar to Fig. 2, except that, in the main part of the Figure, we have added one additional solid thin black line with an arrowhead: the tip of the arrow illustrates the location of a non-magnetic main sequence model of mass 0.095 $M_\odot$ shifts when a magnetic field of a certain strength $B_v$ is included in the stellar model. In the case of the model illustrated by the arrow, we have modelled magneto-convection according to the GT criterion (see eq. 1). We stress that the black arrow in the main part of Fig. 3 refers to a model in which the only difference from the non-magnetic main sequence model of mass 0.095 $M_\odot$ is the inclusion of a non-zero value of $B_v$: there are *no* alterations to $Y$ or to $Z$.

Now that we have determined the shifts in the main sequence location by changing *one* parameter (either $Y$ or $Z$ or $B_v$), we can also determine how simultaneous changes in *more than one* parameter shift the main sequence location. The insert in Fig. 3 shows two possible routes for shifting the 0.095 $M_\odot$ main sequence model so that our model results lie in the empirical error box for Tr-1. Along the dashed line labelled Route (1), $Z$ is kept fixed at $Z = 0.0153$ but $Y$ is increased to $Y = 0.2796$ (as shown by the dashed brown arrow in the insert): in this case, the value of $B_v$ required in order to reach the empirical rectangle is relatively large, as illustrated by the longer (thick) black line in the insert which starts at the tip of the brown arrow and ends inside the empirical rectangle. Along the dashed line labelled Route (2), $Y$ is fixed at $Y = 0.2526$ but $Z$ is increased to $Z = 0.0296$: in this case, the dashed purple arrow (due to enhanced $Z$) moves in the direction of the empirical rectangle, in such a way that when we include magneto-convection a shorter black arrow, when vectorially added to the purple arrow, enables us to reach the empirical rectangle. The shorter length of the black arrow in Route (2) indicates that a weaker magnetic field will fit the empirical rectangle than in Route (1). In general, when we construct magnetic models with a chosen value of Z, the value we obtain for $B_v$ should be considered as an upper limit on the field strength if the star is actually more metal-rich than we assume in our models. Conversely, if the star is actually more metal-poor than we assume in our models, then our result for $B_v$ can be considered as setting a lower limit on the field strength.

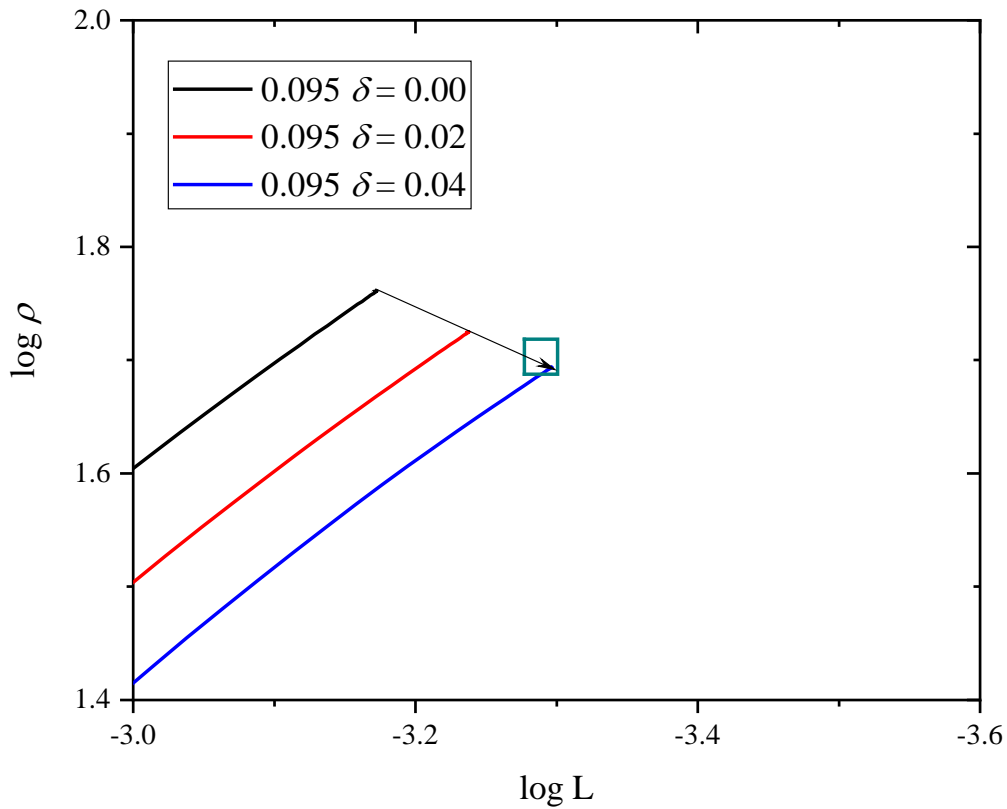

Figure 4. Preliminary evolutionary tracks of magnetic models in the luminosity – mean density plane. The uppermost continuous track refers to a non-magnetic model with mass 0.095 M$_\odot$ : this line ends when the star reaches the main sequence. In this case, the magnetic inhibition parameter δ has a value of zero. With increasing magnetic field strength, the main sequence location shifts to the lower right, along the thin black line with an arrowhead. The tip of the arrow shows the main sequence location when the magnetic field is such that δ = 0.04 in the stellar photosphere. Also plotted is the evolutionary track of a magnetic model with δ = 0.02. The location of Tr-1 is shown by the small dark rectangle. In this figure, the term "preliminary" means that we use the GT criterion (eq. (1)), i.e. the electrical conductivity is assumed to be infinite. In this limit, the fields are maximally effective for impeding the onset of convection.

Alternatively, agreement with the observations can also be obtained for the Caffau et al. (2011) composition models by including only the magnetic effects if the mass of the model is ~ 0.097 $M_\odot$ (results for this case are not shown).

In order to calculate our magnetic models for Tr-1, in line with our recent modeling (MacDonald & Mullan 2017b), we set the ceiling on field strength to be $B_{ceil}$ = 10 kG. This choice of $B_{ceil}$ is based on the maximum magnetic field strength found in simulations of dynamos in low mass stars (Browning 2008; Yadav et al. 2015). We show in Figure 4 a preliminary set of our magneto-convective evolutionary tracks in the luminosity – mean density plane. In Fig. 4, all tracks have mass 0.095 $M_\odot$ and composition $Y$ = 0.2796, and $Z$ = 0.0153: the differences between tracks are due to differences in magnetic field strength, as shown by the different numerical values of the magnetic inhibition parameter $\delta$ (see eq. 1). In this preliminary set of tracks, we chose *not* to include the effects of finite electrical conductivity in our magneto-convective models: the magnetic fields are assumed to be perfectly coupled to the gas, and therefore maximally effective for impeding the onset of convection. Despite the preliminary nature of the tracks, these models show clearly that magnetic effects lead to inflated radii, and therefore could be relevant in our search for a model that fits Tr-1.

For consistency with all of the observational data (luminosity, mean density, and age), the results in Figure 4 show that the magnetic inhibition parameter $\delta$ (see eq. (1) above) must have a value which lies in the range $\delta$ = 0.03-0.04. Combining these numerical values for $\delta$ with the gas pressure in the photosphere of each model, we find that the corresponding vertical magnetic field strengths $B_v$ in the photosphere of Tr-1 would be $B_{ps}$ ~ 1100 - 1300 G.

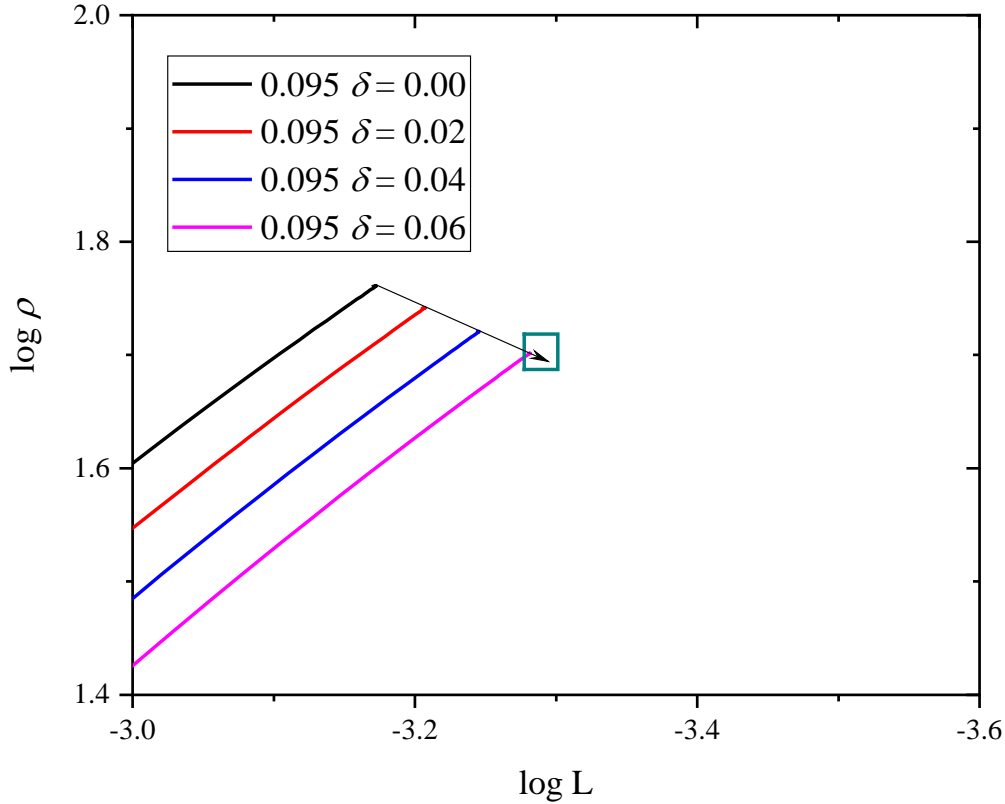

Figure 5. As in Figure 4, except that here, the effects of finite electrical conductivity on magneto-convection have been included in the models.

    The assumption of infinite electrical conductivity which was used in calculating the preliminary tracks shown in Figure 4 cannot be justified in Tr-1. Two lines of argument can be presented to indicate that effects of finite conductivity play a role in stars which are close to (or below) the lower end of the main sequence. First, the slope of the flare frequency distribution changes significantly among the coolest flare stars (Mullan & Paudel 2018): the change in slope can be attributed quantitatively to onset of finite conductivity. Second, MacDonald et al. (2018) have demonstrated quantitatively that at the low temperatures ($T_{eff}$ ~ 2700 – 2800 K) which exist in the surface layers of GJ 65A and GJ65B, the presence of finite electrical conductivity allows the mostly un-ionized gas to drift significantly through the magnetic field lines. That is, the magnetic field does not have "as much of a hold on" the gas as it would have in a medium where electrical conductivity is infinite. In such cases, in calculating a magneto-convective model, we do not rely on the GT criterion, but on the GTC criterion (see Section 3.3 above). Since the effective temperature of Tr-1, $T_{eff}$ ~ 2500 K, is even lower than those of GJ 65A/B, we expect that the effects of finite electrical conductivity are even more significant for Tr-1. Therefore, in order to replicate a given empirical inflation of the radius, the field strength in Tr-1 needs to be larger than it would be if electrical conductivity were infinite. To quantify this in the context of Tr-1, we show in Figure 5 (which has the same format as in Figure 4) the values of $\delta$ which are needed in order to be consistent with the empirical error rectangle. In this case, we now find that the magnetic inhibition parameter must have larger numerical values, namely in the range $\delta$ = 0.05-

0.07. Combining these values with the gas pressure in the photosphere (~$2\times10^6$ dyn cm$^{-2}$), we find that the corresponding vertical field strengths in the photosphere of Tr-1 are $B_{ps}$ ~ 1450 - 1700 G. As expected, these fields are stronger than the fields that would suffice to replicate the empirical inflation (in an idealized model) if the electrical conductivity were in effect infinite (see Figure 4).

## 4. Magnetic fields in Tr-1: poloidal and toroidal components

It is important to note that, when we use the GT (or GTC) criteria to obtain a best-fit magneto-convective model of a star, the principal result that emerges is a numerical value of the parameter $\delta$ at the surface of the star. A key aspect of $\delta$ is that it depends on a particular component of the magnetic field, namely, the *vertical* component $B_v$ of the field which threads through a convection cell. Therefore, the values of $B_v$ = 1450-1700 G obtained above for Tr-1 refer to an average *vertical* field strength over the surface of the star. How should we interpret these field strengths in the context of the global field distribution of magnetic field that exists in the near-surface layers of a low-mass star such as Tr-1? To address this, it is convenient to consider how any divergence-free vector quantity (e.g. a magnetic field) can usefully be decomposed in a spherical system into two components: poloidal and toroidal.

*4.1. Poloidal fields*

In the Sun, the poloidal component includes a global dipole, which is most easily observed in and near the polar regions. This component can be identified in coronal images at all stages of the solar cycle, although it is easier to identify at solar minimum. For example, see Habbal et al. (2010: their Figure 1), where remarkable images are reported of an eclipse which occurred during the deep solar minimum of 2008. Near the poles of the Sun, the field lines on the solar surface are essentially vertical, i.e. radial: $B_{ps} \approx B_v \approx B_r$. Because of this, attempts to derive the Sun's surface field at the poles by observations from Earth are hampered by serious foreshortening. Instead, *in situ* measurements of the radial field in interplanetary space $B_{r,i}$ can be made in a solar wind stream which emerges from a coronal hole. Conservation of magnetic flux requires that, if the *in situ* interplanetary stream extends over an area $A_i$ on a sun-centered sphere, then $B_{r,i} A_i = B_{ss} A_b$ where $A_b$ is the surface area at the base of the coronal hole. In a study of several large solar wind streams observed during the Skylab mission, Hundhausen (1977) reports that $B_{ps}$ ranges from 6 to 12 G.

Poloidal fields need not consist only of dipoles. The global dipole is the lowest order multipole of the magnetic field, with a strength that decreases radially as $r^{-3}$. Higher order multipoles of the poloidal field have strengths that decrease faster with radius. As such, the global dipole retains relatively large values out to farther radial distances from the Sun than any higher order multipole. Therefore, at relatively large altitudes above the surface of the Sun, the corona is threaded predominantly by the global dipole field, with field lines which follow local meridians from one pole to the other. (This global dipole field approximation breaks down beyond a few solar radii: at a certain radial location [the "source surface" of the solar wind], the ram pressure of the solar wind forces the field to open up into a radial topology. However, the "source surface" lies on average at a radial location of about 3.25 $R_\odot$ [Koskela et al. 2017], which is far outside the regions where flares occur. Therefore, in the context of flares, we confine our discussion to the global dipole.)

*4.2. Toroidal fields*

The toroidal component gives rise to active regions and spots, which are confined to temporary localized regions in lower latitudes: in the Sun, these regions emerge from below the surface, develop into more or less large areas, and eventually decay, on time scales of days or weeks. The strongest values of toroidal fields in the photosphere are observed in sunspot umbrae, where fields up to 2 kG are present in 80% of spots, while 5% of spots have fields exceeding 3 kG (BL64, p. 206). According to the Babcock (1961) dynamo model, the strong toroidal fields which give rise to sunspots are derived from the poloidal field by means of differential rotation which stretches the poloidal field lines (initially confined to N-S planes) into lines which become progressively aligned in the E-W direction. How strong can the toroidal fields become in principle? It is difficult to see how the fields can become much stronger than the limit which is set by the gas pressure $p_g$ in the photosphere: if the magnetic pressure $p_m = B^2/4\pi$ were to exceed $p_g$, a magnetic flux rope would expand horizontally, weakening the value of $B$ until (rough) equipartition is established between $p_m$ and $p_g$. In fact, given $p_g \sim 1.2 \times 10^5$ dyn/cm$^2$ (e.g. Mullan 2009), the equipartition field is predicted to be 1.7-1.8 kG, consistent with the observed fields in the umbrae of most sunspots.

*4.3 Which component of the field is involved in solar flares?*

When we wish to consider the energy that is released in flares, the strength of the magnetic field in the *photosphere* is not directly pertinent. The gas in the photosphere is mainly electrically neutral (with a degree of ionization of order $10^{-4}$ [Mullan 2009, p. 59]): such a gas is not capable of carrying significant electrical currents. Instead, in the context of flares we need to consider the magnetic fields in *the corona*, where the highly ionized plasma allows strong currents to flow. Instabilities in those currents, possibly leading to magnetic reconnection, are believed to be the ultimate source of flare energy (e.g. Alfven & Carlqvist 1967; Hood & Priest 1979). The toroidal component of the Sun's field gives rise to coronal loops that rise above the surface of an active region: the field strengths in coronal loops can be obtained from observations in X-rays and radio, indicating coronal field strengths ranging from 30 to 583 G (Schmelz et al. 1984). If a flare occurs in a solar active region, it is these coronal fields that ultimately provide the energy. In any particular active region, the foot-points of a loop are subject to jostling by convective cells (granules) in the photosphere, and these jostlings can lead to twisting of the field lines. If the twist exceeds a critical value, the field becomes unstable, and a flare starts in the corona (Mullan and Paudel 2018). The energy $E$ which is released in a flare depends on the volume of the loop $V$ and the change in magnetic energy ($\sim B^2$) in the coronal part of the loop: $E \approx V \Delta B^2/8\pi$.

At this point, we ask a question which at first sight might seem to be pointless: however, it may set the stage for a subsequent discussion of Tr-1. Does the *poloidal* field in the Sun play a role in solar flares? The answer is almost certainly "No": the energy density of the poloidal field (of order $\leq 1$ erg cm$^{-3}$ on the solar surface at N and S poles) could in principle power a large solar flare ($E \approx 10^{32}$ ergs) only if the flare occupied an enormous volume, of order $\geq 10^{32}$ cm$^3$. Such a volume would have to encompass a significant fraction of the entire material contents of the corona, which are contained in a volume $V(cor) = 4\pi R^2 H \approx 4\times 10^{32}$ cm$^3$ (where $H$ is the density scale height $\approx 7 \times 10^9$ cm in coronal gas with $T = 10^6$ K). In fact, this estimate of $V(cor)$ is much larger (by a factor of order $10^3$) than the volume of the largest solar flare reported by Aschwanden et al. (2015). For this reason, little attention has been paid to considering flares in the poloidal field of the Sun. By contrast, toroidal fields in the Sun can readily supply flare energies even when confined to a limited volume in a single active region.

*4.4. Interpreting our magnetic models of Tr-1: poloidal or toroidal?*

In order to interpret the numerical results of our magneto-convective modeling of a particular star in a realistic manner, we need to determine which component of the magnetic field (poloidal, toroidal) is more relevant in determining the global properties of that star. The toroidal fields are certainly stronger than the poloidal field, but the toroidal fields are confined to only certain areas of the Sun, and they last only for a finite period of time. Moreover, relative to the surface of the Sun, most of the toroidal fields are horizontal: such fields are much less efficient at interfering with convection than vertical magnetic fields. (A demonstration of this statement can be seen in sunspots, where vertical field lines lead to a cooling of the umbra relative to the photosphere by 1625 K [BL64, p. 107], whereas the horizontal field lines in the penumbra lead to a cooling of only 285 K [BL64, p. 145]: thus, vertical fields generate 5-6 times more cooling relative to the photosphere than horizontal fields do.) On the other hand, over a large fraction of north and south hemispheres, the poloidal field is mainly vertical and is present at essentially all times (apart from a short time near solar maximum when the poloidal field switches polarity). This leads us to the following hypothesis: the poloidal field is more effective at impeding the onset of convection on a global scale than the toroidal field is.

Can this hypothesis be tested? Yes: in the case of the Sun, we can test it by considering two empirical properties: (i) the frequencies of p-modes, and (ii) the radius of the Sun. As regards (i), Mullan et al. (2007) have used magneto-convective modeling to determine how the frequencies of p-modes change as the Sun goes from solar minimum to solar maximum. Given the poloidal field which is known to have values in the range 6-12 G (Hundhausen 1977), and given the photospheric gas pressure of order $1.2\times 10^5$ dyn cm$^{-2}$ (e.g. Mullan 2009), the numerical value of $\delta$ in the polar regions of the Sun's surface is of order $10^{-5}$. Models of the Sun with this value of $\delta$ allow us to calculate the amount by which the p-mode frequencies alter between solar minimum and maximum (Mullan et al. 2007): the model predicts that p-modes with frequencies of 2000-4000 microHz should shift to higher frequencies at solar maximum by amounts which are several tenths of one microHz. These predictions are consistent with empirical frequency shifts. As regards (ii), our solar models predict that the change in field between solar minimum and maximum should lead to a fractional change in the radius of the Sun of $\Delta R/R \approx (0.7-3)\times 10^{-5}$ with a smaller radius at solar maximum. Inserting $R = 7\times 10^5$ km, this predicts that at solar maximum, the solar radius should be smaller by 5-20 km: this is consistent with a report (Kosovichev & Rozelot 2018) that the "seismic radius" of the Sun decreases by 5-8 km at solar maximum.

In view of these checks on our magneto-convective model as applied to the Sun, we consider that it is appropriate to regard the parameter $\delta$ in our Tr-1 model as referring to the *poloidal* field strength near the poles. As a result, we consider that the fields of 1450-1700 G, which we have derived from magneto-convective models for Tr-1 are best interpreted as referring to a global dipole field.

Is it possible that our suggestion of a global field of 1450-1700 G on Tr-1 might be inconsistent with the average field strength (600 G) reported by Reiners & Basri (2010) in their observational study of Tr-1? Not necessarily: we may cite the example of another star with spectral type close to that of Tr-1: the M6.5 – M7 star GJ 3622 = 2MASS J10481258-1120082. For this star, Reiners & Basri report $Bf$ = 600 +/- 200 G. However, for the very same star, Shulyak et al. (2017) report a field strength of $<B>$ = 1400 +/- 200 G: the discrepancy is possibly due to the use of different line fitting methods. If a similar discrepancy could be shown to hold for Tr-1, perhaps a measurement of the field of Tr-1 using the approach of Shulyak et al (2017) might also result in an increase in field strength from 600 G (Reiner & Basri 2010) to 1400 G. It is also important to note that, in their modeling of the line profiles, Shulyak et al. (2017) assume that the magnetic field is dominated by its *radial* (i.e. vertical) component: this is the component which, according to our mode, has a strength of 1450 - 1700 G in order to replicate the empirical radius of Tr-1.

*4.5. Dynamo action in Tr-1*

If our interpretation of $B_{ps}$ ≈ 1450-1700 G as the global dipole field on Tr-1 is reliable, it implies a striking contrast between the strength of the polar field of the Sun (6-12 G) and the strength of the polar field of Tr-1: the latter exceeds that in the Sun by a factor $\varphi$ of order 100-300. Interestingly, this range overlaps with the prediction in Section 2 (para. 1) above, where, based on Kippenhahn's scaling between dynamo field strength and the angular velocity of the star, we predicted $\varphi$ could have a value of as much as 160.

The Babcock (1961) dynamo model which was developed as an explanation for the solar cycle, but it may also operate in other stars. Let us consider an application to Tr-1. Given that the *poloidal* field in Tr-1 is 100-300 times stronger than that in the Sun, should we then conclude that dynamo operation will lead to *toroidal* fields in Tr-1 which are also 100-300 times stronger than the toroidal fields in the Sun? If the answer were yes, it would lead us to conclude that, instead of field strengths of several kG in sunspot umbrae (as in the Sun), the spots on Tr-1 might have umbral field strengths of up to $10^6$ G. Such fields would be greatly in excess of the equipartition field strengths in the photosphere of Tr-1, and therefore seem unlikely. In fact, given that the photospheric gas pressures in Tr-1 are ~$2\times10^6$ dyn cm$^{-2}$, the equipartition fields are expected to be of order 7 kG: remarkably, this is close to the maximum value of photospheric field strengths which have been reported for low-mass stars (Kochukhov & Lavail 2017, Shulyak et al. 2017; Berdyugina et al. 2017). Thus, although in the Sun the differential rotation leads to a toroidal field which can be stronger than the poloidal field by a factor $\theta$ ~ several hundred, our results suggest that this simply cannot be the case in Tr-1: the toroidal field in Tr-1 is limited to a maximum strength that is larger than the poloidal by a factor of no more than $\theta$ ~ 4-5. It is possible that the length of 11-year activity cycle in the Sun is determined by the time required for differential rotation to enhance the field strength by a factor $\theta$ ~ several hundreds (Mullan 2009, pp. 275-279). If this possibility applies also to Tr-1, then the activity cycle could be considerably shorter in Tr-1 than in the Sun (if the differential rotation is comparable).

**5. Magnetic fields on the surface of Trappist-1: consequences for flares**

Now that our models have yielded an estimate of the field strength on the surface of Tr-1, we turn to a consideration of how these results are pertinent to the analysis of the flares that have been detected by the Kepler spacecraft.

Flare data for Trappist-1 from Kepler have been analyzed by Vida et al. (2017) and by Paudel et al. (2018a). The energies of individual flares have been quantified by comparing the flare light curves to the quiescent flux. It appears that, at a given flare frequency, the flare energies obtained by Vida et al differ from those of Paudel et al by factors of 2-3. Vida et al. assume that the quiescent flux follows a blackbody curve with temperature 2550 K, whereas Paudel et al. rely on more empirical spectra for the quiescent flux. In the latter case, strong molecular absorptions in the visible band reduce the stellar flux at those wavelengths below the black body curve by factors of 2-3. As a result, it is not surprising to find that the flare energies estimated by Paudel et al. are smaller by factors of 2-3 than those obtained by Vida et al.

According to Paudel et al. (2018a), flares with optical energies up to a maximum value of $(2-3)\times 10^{32}$ erg have been detected on Tr-1. Since flare energy is related to a combination of flare volume and change in magnetic energy, $E \approx V \Delta B^2/8\pi$, we can ask: is it possible to estimate separately the values of $\Delta B^2$ and $V$ in flares?

Of course, in the case of the Sun, separate estimates of $B$ and $V$ are possible: imaging data allow us to determine $V$, and $B$ can be estimated by combining optical and radio data. (The value of $B$ sets an upper limit on $\Delta B^2$.) But in Tr-1, this separation is not possible. However, as it happens, the largest *solar* flares among a sample of 400 events detected by the Solar Dynamics Observatory (SDO) were recorded to have thermal energies of up $(2-3)\times 10^{32}$ erg (Aschwanden et al. 2015), coincidentally equal to the maximum flare energy reported by Paudel et al. (2018a) in Tr-1. Might this coincidence in largest flare energy imply that similar processes are at work in solar flares as in flares on Tr-1? Not necessarily. The maximum linear extent of solar flares is observed to be $L \approx 4\times 10^9$ cm (Aschwanden et al. 2015), i.e. only about 6% of the solar radius. Flares on the Sun are confined to the active regions in which they occur, and even the largest active region on the Sun occupies only a small fraction of the solar surface. The associated maximum solar flare volume ($V \approx 10^{29}$ cm$^3 \approx L^3$: Aschwanden et al. 2015) is less than 0.1% of the solar coronal volume. In this small coronal volume of a solar flare, where the local magnetic field is at most 300-400 G (Aschwanden et al. 2014), the maximum available magnetic energy (assuming the extreme case where all magnetic energy is converted into flare energy, i.e. $\Delta B^2 = B^2$) is $E \approx V B^2/8\pi$. Inserting values, this leads to upper limits on flare energy of $(4\text{-}6) \times 10^{32}$ erg: these are indeed adequate to power the observed thermal energies of the largest SDO flares.

Suppose we were to apply the solar analogy to flares on Tr-1: if the linear extent of a flare on Tr-1 were to be confined to only 6% of the stellar radius (= 0.121 $R_\odot \approx 8.5 \times 10^9$ cm for Tr-1), then the flare would have a linear extent $L \approx 5 \times 10^8$ cm, and a volume of order $L^3 \approx 10^{26}$ cm$^3$. To power a flare with energy $(2\text{-}3) \times 10^{32}$ erg (such as the largest flare reported by Paudel et al. 2018a) from such a volume, the field in the flaring volume, even in the extreme case $\Delta B^2 = B^2$, would have to be as strong as 7 kG. We have already seen (Section 4) that fields of this magnitude may very well exist in the *photosphere* of Tr-1, but such fields are not relevant for flares: only in the *corona* are densities low enough to be conducive to the onset of the instability which can lead to the process of reconnection.

How strong might the coronal fields be in Tr-1? Our results (Section 3 above) suggest that Tr-1 has a global dipole field with a surface strength of 1.45-1.7 kG at the poles: such a field, falling off radially according to $r^{-3}$, would have a strength of $B_d \approx 200$ G in the corona at a height $h = 1$ R$_*$ above the surface. The next higher order multipole (quadrupole), falling off as $r^{-4}$, would create coronal fields which are weaker than on the surface by a factor of 16 at height $h$: even if the surface strength had its extreme magnitude (7 kG), the quadrupole field $B_q$ would contribute no more than $\approx 400$ G to the field at coronal height $h = 1$ R$_*$ above the surface. More likely, the higher multipole would contribute less than 400 G, perhaps 200 G. In such a case, if the lowest and next lowest multipoles contribute fields which are randomly oriented relative to each other, the total field strength $B_t$ at a height $h = 1$ R$_*$ above the surface could be of order $B_t \approx \sqrt{(B_d^2 + B_q^2)}$ $\approx 300$ G. Higher order multipoles would contribute weaker fields at height $h$. Thus, at coronal height $h$, the poloidal field *might* contribute magnetic energy in the possible flare volume that would be comparable to that contributed by the toroidal field. With a total coronal field of order 300 G, the volume needed to power a flare with an energy of $(2\text{-}3) \times 10^{32}$ erg would be of order $8 \times 10^{28}$ cm$^3$. The linear extent of such a volume would be of order $4\text{-}5 \times 10^9$ cm. This is a sizeable fraction (50%) of the stellar radius of Tr-1, and is much larger than the solar analogy estimated above (where the linear extent amounted to only 6% of the radius). If these estimates have any relevance to stellar flares, they suggest that whereas solar flares are confined to volumes that are small compared to the volume of the parent star itself, this may not be true in the case of the largest flares on Tr-1.

If the poloidal fields do indeed contribute to the largest flares on a low-mass star, whereas toroidal fields dominate in the smaller flares, then the flare frequency distributions for low-mass stars might obey different laws: one power law for the toroidal flares, and another for the poloidal flares. It remains to be seen whether this proposal has any observational signature.

**6. Global field structure in stars of different spectral types**

The discussion in Section 5 above was based on an attempt to infer certain features of the global field in Tr-1 using the properties of the global field in the Sun for guidance. The conclusions which were drawn in Section 5 are therefore expected to be reliable to the extent that we may consider the global topology of a G2 dwarf and that of an M8 dwarf as being (at least somewhat) comparable.

At first sight, it might appear rather unlikely that comparable topologies would exist in such structurally different stars. After all, the dynamo in the Sun almost certainly relies on an αΩ dynamo in the tachocline (between radiative core and convective envelope), whereas Tr-1 (which is completely convective) cannot have an αΩ dynamo, and must rely on a different mechanism to generate its magnetic fields, possibly a turbulent ($\alpha^2$) dynamo (Durney et al 1993).

If the distinction between stars with or without a tachocline carried over in a one-to-one manner to dynamo operation, we might expect to observe a change in coronal emission (which is magnetically driven) at the transition to complete convection (TTCC at spectral type M3 [Houdebine and Mullan 2015; Houdebine et al 2017]). We might also expect to observe a clear change in magnetic topology at the TTCC as well. However, the observations do not confirm these expectations. First, there is no observable change in coronal emission intensity at spectral type M3 (Fleming et al. 1993). Second, cool dwarfs exhibit a variety of magnetic topologies (Morin et al. 2010): some have fields that are strong and axisymmetric (i.e. poloidal), others have weak fields that are not axisymmetric, while still others have fields that include a toroidal component. In a remarkable plot (their Fig. 15), Morin et al. demonstrate that among completely convective stars, one group exhibits a high degree of axisymmetry in their fields, while another group of stars (also completely convective) have fields which are non-axisymmetric: there are roughly equal numbers of stars in both groups. Also among stars with radiative cores, Morin et al. again find a mixture of stars with axisymmetric and non-axisymmetric topology, again with roughly equal numbers in both groups. The data suggest that dynamo operation in cool stars leads to a mixture of field topologies whether the star is completely convective or the star has a radiative core: the mixture of topologies might be due to bistability (Gastine et al. 2013), or might occur because the dynamo enters an oscillatory mode (Kitchatinov et al. 2014).

It would help if theories of the dynamo could present clear-cut guidance as to how best to proceed in interpreting the stellar data. Unfortunately, theoretical dynamo models contain enough complexities that they lead to a bewildering array of possibilities. E.g., in the Babcock (1961) solar dynamo, an essential role is played by differential rotation in the surface layers of the Sun. Inclusion of differential rotation in a dynamo model for a completely convective star led Dobler et al. (2006) to conclude that the dynamo-generated fields in stars would have the property of axisymmetry. However, differential rotation of solar-type need not exist in all cool stars: in fact, in the M3.5 dwarf V374 Peg, evidence suggests that there is *no* differential rotation on the surface (Morin et al. 2008; Vida et al. 2016). In view of this, rigid-body rotation was included in dynamo models computed by Kueker & Ruediger (1999) and also in those computed by Yadav et al. (2015): in the former, an $\alpha^2$ dynamo generated non-axisymmetric fields (resembling a tilted dipole), whereas in the latter, the dynamo generated a dipole-like (i.e. axisymmetric) field.

As regards the possibility of an oscillatory dynamo in any particular star, we would expect to see an activity cycle in that star: certainly the Sun undergoes such a cycle, with a period of roughly 11 years. Is there evidence that low-mass stars have activity cycles? Perhaps: **Savanov (2012), Vida et al (2013), and** Reinhold et al. (2017) may be cited in this regard. **Savanov (2012) analyzed data for 31 M dwarfs spanning intervals of 2000-3000 days: activity cycles were reported with periods ranging from less than 1 year to almost 10 years. Interestingly, the cycle periods were found to be independent of rotational period: the author noted that this conclusion "differs significantly" from the results of earlier studies of non-M dwarfs, perhaps indicative of different dynamo modes. Vida et al (2013) reported on 4 stars with fast rotation, with spectral type ranging from K3 to M4: three of the stars showed activity cycles with periods of 1-2 years, but the M4 star in their sample showed no evidence of periodicity. Vida et al claimed that the absence of long-term cyclic behavior in their M4 star "is consistent with the properties of an $α^2$ dynamo" at work in that completely convective star. However, in at least one other completely convective star, Savonov (2012) reported that periodic activity was in fact present.** Reinhold et al (2017) report that 3000 of the 100,000+ stars in the original Kepler field undergo variations which could be due to cycles with periods in the range 0.5-6 years: **to be sure, Reinhold et al do not draw any particular attention to the M dwarfs, so it is not yet clear how much overlap there is between their results and stars with $T_{eff}$ in the range of interest to us here.** As far as we know, Tr-1 has not been proven to exhibit an activity cycle: so we cannot yet say whether or not Tr-1 shares this characteristic with the Sun.

In view of the empirical mixture of magnetic topologies among stars of comparable mass, it is difficult at the present time to identify any clear correlations between the magnetic properties of stars and their location relative to the TTCC. From this perspective, despite the significant differences in internal structure between the Sun and Tr-1, dynamo activity apparently leads to magnetic topologies which, from an empirical standpoint, may not differ so much that we are definitively precluded from making legitimate comparisons.

In one regard, however, the results of the present paper may be considered to receive some support from the work of Morin et al (2010). Inspection of Fig. 15 in Morin et al. (2010), suggests that the stars with the strongest fields (with magnetic fluxes of up to 1-2 kG) lie preferentially *below* the TTCC (i.e. later than M3), while the fields in general become weaker as we examine stars *above* the TTCC, i.e. stars with masses approaching the mass of the Sun (where magnetic fluxes are of order tens of G). This feature is consistent with our conclusion in the present paper that the polar field (1-2 kG) on the M8 dwarf Tr-1 is significantly stronger (by ~100) than the polar field in the Sun.

**7. Is a planet in the HZ of Tr-1 unsuited for life? CME effects**

Sagan (1973) proposed that flares are harmful to life because of UV photons: we will consider Sagan's argument in Section 7. In this section, we consider a different hazard that could also adversely affect life on exoplanets in orbit around a flare star. Coronal mass ejections (CME's) are a well-studied aspect of magnetic activity in the Sun: in the years 1996-2010, the SOHO spacecraft recorded 12,433 CME's (Youssef 2012), i.e. one CME in about 50% of all flares. CME's of sufficient energy may strip the atmosphere from an unprotected (i.e. unmagnetized) planet, so it is relevant to ask: how effective are flare stars at generating CME's which actually put a planet in peril as regards life?

The answer is not yet clear. The occurrence of a CME-like phenomenon has been reported in at least one flare star (the cool component of V471 Tau) with a spectral type of K2 (e.g, Bond et al. 2001): such a star is similar to the Sun in showing evidence that flares occur in the vicinity of spots (Young et al. 1983). **As regards CME's in an M dwarf, Vida et al (2016) reported on their detection of moving material reminiscent of a CME in the M4 star V374 Peg: however, the motions involved material falling back to the star, which led Vida et al (2016) to label two events as "failed CME's", while a third event was considered a "real" CME. They reported a CME-rate that is "much lower than expected from extrapolations of the solar flare-CME relation". Moreover, a variety of observational evidence indicates that CME events in M dwarfs with material clearly moving faster than escape speed are rare. Thus, Leitzinger et al (2014) used Hα to search for flares and CME's in 28 dK-dM stars in a young cluster: although flares were observed, they detected "no distinct indication" of CME's. And Korhonen et al (2017) examined archival Hα spectra for 40 single active stars, but found no CME's except in the one event reported by Vida et al (2016). Korhonen et al (2017) also observed G-K-M stars in 5 clusters with a range of ages: in 4 clusters, some flares were detected, but no CME's. In one cluster, one early M star showed variability in Hα which might be related to a CME: in conclusion, they wonder "Maybe we are not detecting many CME's because there actually are only few of them?" As regards radio evidence for the relative rarity of CME's from flare stars, we note that Crosley and Osten (2018) conducted** a 64-hour long search to search for evidence of CME's ejected by an active flare star (EQ Peg: a binary with spectral types M4 and M5). Relying on evidence of Type II radio bursts (drifting from high to low frequencies as a shock wave propagates outward through the corona), the search failed to yield evidence of such features. The authors concluded that their result "casts serious doubt on the assumption that a high flaring rate [in an M dwarf] corresponds to a high rate of CMEs". **Moreover, Villadsen and Hallinan (2018), who obtained radio data for 5 active M dwarfs over 58 hours and detected frequent bursts of coherent emission, found no evidence for analogs of Type II bursts "which are often driven by super-Alfvenic CME's". Thus, the observed frequency of stellar CME's is apparently not consistent with naively scaling from the solar case.**

Based on the results we have obtained in the present paper, we would like to point out a possible reason for the lack of CME's in an active flare star (such as EQ Peg, or Tr-1): 3-D MHD modelling of CME's has shown (Alvarado-Gomez et al. 2018: hereafter AG18) that if a star has a "strong enough" global dipole field, a CME with a certain kinetic energy (KE) originating in the low corona may be trapped by the overlying dipole field, and be unable to escape from the star. Such CME's would pose no peril to a planet in the HZ. How strong is "strong enough"? For the largest solar-like CME's, where the KE has been observed to as large as $\sim 3 \times 10^{32}$ ergs, AG18 find that a global dipole with a critical strength of $B_c = 75$ G is sufficient to suppress the escape of such CME's. Since flares on Tr-1 have energies which are observed to extend up to $\sim 3 \times 10^{32}$ ergs (Paudel et al. 2018a), CME's of comparable energy might be associated with the observed flares. In such a case, it does not seem implausible to expect that the suppression of CME escape reported by AG18 for the Sun could also apply to all of the flares which have (so far) been *observed* on Tr-1 provided that the global dipole in Tr-1 had a strength of at least 75 G. This is where the results of our magneto-convective models come into play: our modeling in this paper indicates that the global dipole field of 1450-1700 G on Tr-1: these field strengths are well in excess of the 75 G limit discussed by AG18. Therefore, the fields we have derived for Tr-1 can be expected to have the capacity to suppress readily any "solar-like" CME's on Tr-1 with energies up to the largest values which occur in *observed* flares.

To be sure, the AG18 models also consider the possibility that a star may occasionally generate CME's with KE larger than 3 x $10^{32}$ ergs: such events (which they label as "monster" CME's) are assigned a KE as large as $10^{34-35}$ ergs. There is no record that the Sun has experienced flares with energies in excess of $10^{33}$ ergs over the past 2000 years (Maehara et al 2012). CME's of "monster" magnitude have never been reported (so far) in the Sun, nor have flares of such energies been detected in Tr-1 (so far): the maximum flare energy reported by Paudel et al (2018a) is only (2-3) x $10^{32}$ ergs. **It is true that "superflares" have been reported on solar-like stars with energies >$10^{33}$ ergs: Maehara et al (2012) have reported on 365 such events which might possibly enter into the AG18 class of "monster" events. However, among M dwarfs, such as the star we are interested in here, "monster" flares are considerably rarer. For example, Kovari et al (2007) have reported on one flare on an M dwarf with an energy of order $10^{35}$ ergs. Moreover, for a very young brown dwarf, two "monster" releases of optical energy ($10^{36-38}$ ergs) have been reported (Paudel et al 2018b): however, it is possible that the latter events are related to accretion episodes, rather than being magnetic phenomena analogous to *bona fide* solar-like flares. Besides, even if we know the optical energy released in a "flare", the pronounced rarity of CME's among detectable stellar flares (see discussion in the second paragraph of this Section) means that at present we seem to have no reliable method to predict the energy associated with a CME which might (or might not) be ejected in connection with any particular flare.**

However, if such "monster" events were to occur in the Sun, the models of AG18 indicate that the CME would be able to escape even if the global dipole field were to be of order 75 G. How might this result be applied to Tr-1? From an energetic standpoint, even the "monster" events might be unable to escape from Tr-1 *if* the coronal field were to be correspondingly stronger: a scaling in which $B_c^2/8\pi \approx$ KE could be plausible. In such a case, the "monster" events modeled by AG18 (with KE = $10^{34-35}$ ergs) could be suppressed in the presence of an overlying dipole field $B_c \approx 750$ G. A crucial aspect of the present paper is that this estimate of $B_c$ is exceeded by the global dipole fields we have obtained for Tr-1. This leads us to hypothesize that all CME's in Tr-1 (even the "monster" ones) could well be suppressed by the overlying global dipole field which we are suggesting in this paper (1400-1750 G, i.e. $> B_c$). A full modeling effort (which is beyond the limits of the present paper) would be needed to evaluate the correctness of this hypothesis.

To the extent that our hypothesis is correct, we suggest that Tr-1 has a strong enough global field to mitigate (in effect) the CME hazard as regards any life-form that may be struggling to survive on a planet in the HZ of Tr-1.

**8. Is a planet in the HZ of Tr-1 unsuited for life? Photon effects**

Since the work of Sagan (1973), it has been believed that UV photons with wavelengths from 2400 to 2900 Å would be harmful to a living cell, especially during solar flares. The flux of UV photons was estimated to be so strong at 2400-2900 Å during a solar flare that a living cell would be subject to a lethal does in a time interval that would be no longer than 0.3 seconds. If this conclusion were applicable to Tr-1, the chances of living cells surviving on a planet in orbit around Tr-1 would be small. This result has led Vida et al (2017) to the conclusion that, as regards the planets around Tr-1, the words "unsuited for life?" should be included in the title of their paper.

However, Sagan proposed two methods for protecting the cell from the damaging effects of UV photons. One was to place the cell some 10's of meters underwater: such an approach would work on a world provided that water gathers into deep enough ponds (e.g. Pearce et al 2017). But even if there are no ponds on a planet in orbit around Tr-1, Sagan's second suggestion was to use certain organic molecules (especially the bases which occur in genetic material) to absorb the harmful photons in the outer layers of the cell, preventing them from reaching the genetic material in the nucleus. Thus, UV photons are not automatically harmful to cell life.

In fact, under some circumstances, especially in the vicinity of flare stars, the UV photons from flares may actually be beneficial for the origin of life (Ranjan et al. 2017). Moreover, once life has gotten started on planets that are in orbit around M dwarfs, the visible photons from flares also serve to enhance the efficiency of photosynthesis by as much as an order of magnitude (Mullan & Bais 2018).

In view of these points, we suggest that the effects of flares on life are not necessarily always negative. On the contrary, both of the above photonic processes may play a role in enhancing the probability of life on the HZ planets around the flare star Tr-1.

**9. Conclusions**

Using a model of magneto-convection in which the onset of convection in an electrically conducting medium is impeded quantitatively by the presence of a vertical magnetic field, we have obtained a magnetic model of Tr-1 (spectral type M8). Our model includes the effects of finite electrical conductivity, and the results are found to fit all of the available empirical data for Tr-1 (radius, mass, effective temperature, metallicity). Our models predict that the vertical component of the surface magnetic field on Tr-1 is in the range ~ 1450-1700 G. We argue that this field strength applies to the global dipole field in the photosphere at the magnetic poles. If this argument is valid, the surface field on Tr-1 is stronger than the global dipolar field on the Sun (6-12 G: Hundhausen 1977) by factors of ~100. This enhancement in field strength may be related to the fact that the angular velocity of rotation of Tr-1 exceeds that in the Sun by factors of 10-30.

Compared to the recent modeling which we have reported on two other flare stars, GJ65A/B, with spectral types M5.5 and M6 (MacDonald et al. 2018), the surface field on Tr-1 is weaker by a factor of about 2. The weakness of the Tr-1 field compared to GJ65A/B may be related to the slower rotation of Tr-1, which rotates at least 3 times slower than GJ65A/B. However, both systems are rotating with sufficiently short periods (0.3-0.8 days) that the dynamos in both systems may be saturated: if that is the case, we may be seeing simply the intrinsic scatter of a dynamo in the saturation regime, where empirical field strengths are known to exhibit scatters by factors of 2 (Reiners et al. 2009).

Contrary to a question raised in the title of an article by Vida et al. (2017), we suggest that a planet in the HZ of Tr-1 need not be considered "unsuited for life" despite the occurrence of well-documented flaring activity on the parent star. In fact, we believe that conditions near Tr-1 may even be beneficial for life for the following reasons: (i) the presence of flare photons at certain ultraviolet wavelengths can be beneficial to the origin of the bases required for nucleic acids; (ii) the presence of flare photons at visible wavelengths of 400-700 nm can enhance the efficiency of photosynthesis; (iii) the strong global magnetic field which we have modeled on Tr-1 may suppress CME's to the extent that the violent effects of CME's are less likely to pose a threat of removing the atmosphere of the planet.

*Acknowledgements*
We thank an anonymous referee for constructive comments on the paper, especially as regards several valuable citations. JM and HF acknowledge support from the NASA Delaware Space Grant.